\documentstyle[aps,twocolumn,epsfig]{revtex}

\begin{document}
\def\CC{{\rm\kern.24em \vrule width.04em height1.46ex depth-.07ex
\kern-.30em C}}
\def\P{{\rm I\kern-.25em P}}
\def\RR{{\rm
         \vrule width.04em height1.58ex depth-.0ex
         \kern-.04em R}}

\draft

\twocolumn[\hsize\textwidth\columnwidth\hsize\csname@twocolumnfalse\endcsname

\title{Quantum entanglement of unitary  operators on bi-partite systems}
\author{Xiaoguang Wang and Paolo Zanardi}
\address{Institute for Scientific Interchange (ISI) Foundation, Viale Settimio Severo 65,
I-10133 Torino, Italy}
\date{\today}
\maketitle

\begin{abstract}
We study the entanglement of unitary  operators on $d_1\times d_2$ quantum systems.
This quantity is closely related to the entangling power of the associated quantum
evolutions.
The  entanglement of a class of unitary operators is quantified by the 
concept of concurrence. 
\end{abstract}

]
\section{introduction}

Quantum entanglement plays a key role in  quantum information theory. In
recent years, there has been a lot of efforts to characterize the
entanglement of quantum states both qualitatively and quantitatively. 
Entangled states can be generated from disentangled states by the action of
a  non-local Hamitonians. 
That means these Hamiltonians have the ability to
entangle quantum states. 
It is therefore  natural to investigate the entangling
abilities of  non-local Hamiltonians and the corresponding unitary
evolution operators. The first steps along this direction have been performed%
\cite{Zanardi,Dur} recently.

In Ref. \cite{Zanardi} it has been analyzed the entangling capabilities of
unitary operators on a $d_1\times d_2$ systems and
 introduced an  entangling power measure given by the mean linear
entropy produced by acting with the unitary operator on a given distribution
of product states. D\"{u}r {\em et al. }\cite{Dur} investigated the
entanglement capability of an arbitrary two-qubit non-local Hamiltonian and
designed an optimal strategy for entanglement production. Cirac {\em et al. }%
\cite{Cirac} studied which physical operations acting on two spatially
separated systems are capable of producing entanglement and shows how one
can implement certain nonlocal operations if one shares a small amount of
entanglement and is allowed to perform local operations and classical
communications.

The notion of entanglement of quantum evolutions e.g., unitary operators,
has been  introduced in Ref. \cite{Zanardi1} and  there quantified by  
linear entropy\cite{Zanardi1}. As discussed in that  paper, this  notion arises
in a very natural way once one recalls that unitary operators of a multipartite
system belong to a multipartite state space as well, the so-called Hilbert-Schmidit 
space. It follow that one  can lift all the notions developed for entanglement 
of quantum states to that of quantum evolutions. 
In this report we shall make a further step by   studying the entanglement of 
a class of useful unitary operators e.g., quantum gates, on general
 i.e., $d_1\times d_2$ bipartite quantum systems.

\section{operator entanglement}

We shall denote the $d$-dimensional Hilbert state space by ${\cal H}_d.$ The
linear operators over ${\cal H}_d$ also form a $d^2$-dimensional Hilbert
space and the corresponding scalar product between two operators is given by
the Hilbert-Schmidt product $\langle A,B\rangle :=$tr$(A^{\dagger }B),$ and $%
||A||_{HS}=\sqrt{\text{tr}(A^{\dagger }A)}.$ We denote this $d^2$%
-dimensional Hilbert space as ${\cal H}_{d^2}^{HS}$ and the bra-ket
notations will be used for operators. Let $T_{ij}$ be the permutation (swap)
operator between the Hilbert space $H_{d_i}\otimes H_{d_j}$ ($d_i=d_j$) and
denote by $\hat{T}_{ij}$ its adjoint action, i.e., $\hat{T}%
_{ij}(X)=T_{ij}XT_{ij}.$ We also define the projectors $P_{ij}^{\pm
}=2^{-1}(1\pm T_{ij})$ over the totally symmetric (antisymmetric) subspaces
of $H_{d_i}\otimes H_{d_j}.$

In this section, following Ref. \cite{Zanardi}, we shall adopt as 
an entanglement measure of (normalized) unitary $|U\rangle \in 
{\cal H}_{d_1^2}^{HS}\otimes {\cal H}_{d_2^2}^{HS}$ , the linear entropy,

\begin{equation}
E(U)=1-\text{Tr}_1\rho _U^2,\text{ }\rho _U=\text{Tr}_2|U\rangle \langle U|.
\end{equation}
Note that Tr and tr denote the trace over ${\cal H}_{d^2}^{HS}$ and 
${\cal H}_d,$ respectively. Using the identity tr$_{12}[(A\otimes B)T_{12}]=$tr$%
_1(AB),$ one can easily obtain\cite{Zanardi1}

\begin{eqnarray}
E(U) &=&1-\frac 1{d_1^2d_2^2}\text{Tr}_{13}[(\rho _U\otimes \rho _U)\hat{T}%
_{13}]  \nonumber \\
&=&1-\frac 1{d_1^2d_2^2}\langle U^{\otimes 2},\hat{T}_{13}(U^{\otimes
2})\rangle  \nonumber \\
&=&1-\frac 1{d_1^2d_2^2}\text{tr(}U^{\otimes 2}T_{13}U^{\dagger \otimes
2}T_{13}).  \label{eq:e}
\end{eqnarray}
The term $1/(d_1^2d_2^2)$ is nothing but the normalization factor of $%
U^{\otimes 2}.$
Now we a give a relation between the entanglement of unitaries and the
entangling power introduced in Ref.\cite{Zanardi}
that extends the analogous oen given in \cite{Zanardi1}.

The entangling power $%
e_p(U)$ of a unitary $U$ is defined over ${\cal H}^{\otimes 2}$ as an the average
of the entanglement $E(U|\Psi \rangle ),$ where the $|\Psi \rangle ^{\prime }
$s are product states generated according to some given probability
distribution $p.$ By choosing for $E$ the linear entropy and using the
uniform distribution $p_0$, one find\cite{Zanardi}

\begin{eqnarray}
e_{p_0}(U) &=&1-\frac 4{d_1(d_1+1)d_2(d_2+1)}  \nonumber \\
&&\times \text{tr}(U^{\otimes 2}P_{13}^{+}P_{24}^{+}U^{\dagger \otimes
2}T_{13}).  \label{eq:epower}
\end{eqnarray}
By comparing Eqs.(\ref{eq:epower}) and (\ref{eq:e}), after some straightforward
algebra, we find

\begin{eqnarray}
e_{p_0}(U) &=&\frac{d_1d_2}{(d_1+1)(d_2+1)}  \nonumber \\
&&\times \left[ E(U)+\tilde{E}(U)+\frac 1{d_1d_2}-1\right] ,
\label{eq:relation}
\end{eqnarray}
where

\[
\tilde{E}(U)=1-\frac 1{d_1^2d_2^2}\text{tr(}U^{\otimes 2}T_{24}U^{\dagger
\otimes 2}T_{13}). 
\]
Eq.(\ref{eq:relation}) shows that the entangling power of an unitary is
directly related to the entanglement of it. When $d_1=d_2,$ Eq.(\ref
{eq:relation}) reduces to\cite{Zanardi}

\begin{equation}
e_{p_0}(U)=\frac{d^2}{(d+1)^2}\left[ E(U)+E(US)-E(S)\right] ,
\end{equation}
where $S$ is the swap operator.

\section{Concurrence for bi-partite operators}

From now on we will focus on the   class of unitary operators  given by

\begin{equation}
U=\mu A_1\otimes A_2+\nu B_1\otimes B_2,  \label{eq:psi}
\end{equation}
where $A_1$ and $B_1$ are operators on $d_1$-dimensional system and
similarly $A_2$ and $B_2$ are operators on $d_2$-dimensional system with
complex values $\mu $ and $\nu .$ 
Several interesting unitary operators and quantum gates, we will see later on,
 are of the form (\ref{eq:psi}).

After normalization with respect to the Hilbert-Schmidt norm, the unitary
operator $U$ is given by

\begin{equation}
|U\rangle =\tilde{\mu}|A_1\rangle \otimes |A_2\rangle +\tilde{\nu}%
|B_1\rangle \otimes |B_2\rangle ,
\end{equation}
where

\begin{eqnarray}
\tilde{\mu} &=&\mu {\prod_{k=1}^2 d_k^{-1/2} \|A_k\|_{HS}}
\nonumber\\
\tilde{\mu} &=&\mu {\prod_{k=1}^2 d_k^{-1/2} \|B_k\|_{HS}}
\end{eqnarray}
Here the normalized operator corresponding to the Hermitian operator $O$ is
denoted by $|O\rangle ,$ i.e., 
\begin{eqnarray}
|O\rangle  &\equiv & 1/{\|O\|_{HS}}O, \quad 
\langle O|O\rangle  =1.
\end{eqnarray}

The two operators $|A_i\rangle$ and $|B_i\rangle$ $(i=1,2)$ are assumed to be linearly
independent and span a two-dimensional subspace in the Hilbert space ${\cal H%
}_{d_i^2}^{HS}$ . We choose an orthogonal basis $\{|0\rangle _i,|1\rangle
_i\}$ as in Ref.\cite{Mann}, 

\begin{mathletters}
\begin{eqnarray}
|0\rangle _1 &=&|A_1\rangle ,  \nonumber \\
|1\rangle _1 &=&\frac 1{{\cal M}_1}(|B_1\rangle -\langle A_1,B_1\rangle
|A_1\rangle ),  \label{eq:basis1} \\
|0\rangle _2 &=&|B_2\rangle ,  \nonumber \\
|1\rangle _2 &=&\frac 1{{\cal M}_2}|A_2\rangle -\langle B_2,A_2\rangle
|B_2\rangle .  \label{eq:basis2}
\end{eqnarray}
where

\end{mathletters}
\begin{eqnarray}
{\cal M}_k^2 &=&1-
-\frac{|\langle A_k, B_k\rangle|^2}{\|A_k|_{HS}^2\|B_k|_{HS}^2}
\nonumber \\
\text{ }k &=&1,2.
\end{eqnarray}
Using this  basis the entangled state $|U\rangle $ can be rewritten as

\begin{eqnarray}
|U\rangle  &=&\tilde{\mu}|0\rangle _1\otimes (\langle B_2|A_2\rangle
|0\rangle _2+{\cal M}_2|1\rangle _2)  \nonumber \\
&&+\tilde{\nu}(\langle A_1|B_1\rangle |0\rangle _1+{\cal M}_1|1\rangle
_1)\otimes |0\rangle _2.  \label{eq:uuu}
\end{eqnarray}
The above `state' can be considered as a two-qubit state. Therefore it is convenient to use one simple entanglement measure, the concurrence\cite{Con}, to quantify
the entanglement in the state $|U\rangle $ .  The concurrence for a pure
state $|\psi \rangle \,$is defined as 
\begin{equation}
{\cal C}=|\langle \psi |\sigma _y\otimes \sigma _y|\psi ^{*}\rangle |.
\label{eq:ccc}
\end{equation}
Here $\sigma _y=-i(|1\rangle \langle 0|-|0\rangle \langle 1|).$ A direct
calculation shows that the concurrence of the unitary operator $|U\rangle $
is given by

\begin{eqnarray}
{\cal C}(U) &=&2|{\cal M}_1{\cal M}_2\tilde{\mu}\tilde{\nu}|=\nonumber\\
& {2|\mu \nu |}&
\prod_{k=1}^2 d_k^{-1} \sqrt{\|A_k|_{HS}^2\|B_k|_{HS}^2-|\langle A_k, B_k\rangle|^2}
\label{eq:c}
\end{eqnarray}
In the derivation of the above equation we have used Eqs.(\ref{eq:basis1}-\ref
{eq:ccc}).
Equation (\ref{eq:c}) defines a (non-negative) real-valued functional over 
the operatorial family (\ref{eq:psi})
The   relevant properties of $\cal C$ are summarized in the following. 

\begin{itemize}

\item[(a)]
From the Cauchy-Schwarz inequality it follows that
${\cal C}=0$ iff either $A_1=\lambda_1  B_1$ or $A_2=\lambda_2 B_2
(\lambda_k\in\CC.)$
This is just the separable
case. 

\item[(b)]  For arbitrary unitary operator of the form $U_1\otimes U_2,$ $%
{\cal C}[(U_1\otimes U_2)U]={\cal C}[U(U_1\otimes U_2)]={\cal C}(U)$, which
is nothing but the invariance of entanglement measure under the local
unitary transformations. 

\item[(c)] The Hermitian conjugation of the unitary
operator $U$ is given by $U^{\dagger }=\mu ^{*}A_1^{\dagger }\otimes
A_2^{\dagger }+\nu ^{*}B_1^{\dagger }\otimes B_2^{\dagger }.$ From Eq.(\ref
{eq:c}) it is straightforward to check that ${\cal C}(U^{\dagger })={\cal C}%
(U).\,$

\end{itemize}

These claims immediately follows from Eq. (\ref{eq:c}) and the properties
of the Hilbert-Schmidt
scalar product.
Notice also that for the orthogonal case, i.e., 
$\langle A_k, B_k\rangle=0\,(k=1,2),$ Eq.(\ref{eq:c})  reduces to

\begin{equation}
{\cal C}(U)={2|\mu ||\nu |}
\prod_{k=1}^2 d_k^{-1} \|A_k|_{HS}\|B_k\|_{HS}
\label{eq:cc}
\end{equation}
Furthermore  if the operators $A_i$ and $B_i$ are Hermitian and self-inverse,
i.e., $A_i^2=B_i^2=1,$ then Eq.(\ref{eq:cc}) reduces to

\begin{equation}
{\cal C}(U)=2|\mu ||\nu |.  \label{eq:ccccc}
\end{equation}

\section{examples} 
In the following we consider several explicit examples of the unitary operators $U$.

\subsection{ $2\times(2j+1)$}
We consider a unitary operator acting on $2\times d_2$
systems, which is defined as

\begin{equation}
U_{2\times d_2}=e^{-i2\theta \sigma _zJ_z}=I\otimes \cos (2J_z\theta )-i\sigma _z\otimes
\sin (2J_z\theta ),
\end{equation}
where $d_2=2j+1,$ $\sigma_z$ is the Pauli matrix, and $J_z$ is the $z$ component of the spin-$j$ angular
momentum operator $\vec{J}.\,$
This unitary operator describes the interaction between a spin-1/2 and spin-$j$. 
After normalization $U_{2\times d_2}$ is written as
\begin{eqnarray}
|U_{2\times d_2}\rangle  &=&\sqrt{\frac c{2j+1}}|I\rangle \otimes |\cos (2J_z\theta
)\rangle   \nonumber \\
&&-i\sqrt{\frac s{2j+1}}|\sigma _z\rangle \otimes |\sin (2J_z\theta )\rangle , 
\end{eqnarray}
where

\begin{mathletters}
\begin{eqnarray}
c &=&\text{Tr}(\cos ^2(2J_z\theta ))=\frac{2j+1}2+\frac x2, \\
s &=&\text{Tr}(\sin ^2(2J_z\theta ))=\frac{2j+1}2-\frac x2, \\
x &=&\sum_{k=-j}^j\cos (4k\theta )=\frac{\sin [2(2j+1)\theta ]}{\sin
(2\theta )}.
\end{eqnarray}

The concurrence is obtained as

\end{mathletters}
\begin{equation}
{\cal C}(U_{2\times d_2})=\sqrt{1-\frac{\sin ^2[2(2j+1)\theta ]}{%
(2j+1)^2\sin ^2(2\theta )}}.
\end{equation}

In figure 1 we give a plot of the concurrence against $\theta .$ 
The period of the above function with respect to $\theta$ is $\pi/2$.
So we just plot the figure and make discussions within one period.
We see that there is $2j$ maximal points in one period at which operator is maximally entangled. 


\subsection{$(2 j_1+1)\times (2j_2+1)$}

\begin{eqnarray}
U_{d_1\times d_2} &=&e^{-i\pi {\cal N}_1\otimes {\cal N}_2}  \nonumber \\
&=&\frac 12[1+\Pi _1]\otimes I+\frac 12[1-\Pi _1]\otimes \Pi _2,
\label{eq:qqq}
\end{eqnarray}
where ${\cal N}_i=J_{iz}+j_i$ is the number operator, $\Pi _i=(-1)^{{\cal N}%
_i}$ ($i=1,2$) is the parity operator of system $i,$ and $d_i=2j_i+1.$
This operator describes the interaction between spin-$j_1$ and spin-$j_2$ and it can be used to generate entangled SU(2) coherent states\cite{ECS}.

By comparing Eqs. (\ref{eq:psi}) and (\ref{eq:qqq}), we find 
\begin{eqnarray}
\mu  &=&\nu =1\text{ },  \nonumber \\
A_1 &=&\frac 12[1+\Pi _1],B_1=\frac 12[1-\Pi _1],  \nonumber \\
A_2 &=&I,B_2=\Pi _2,  \label{eq:qqq1}
\end{eqnarray}
which have following properties

\begin{eqnarray}
A_1^2 &=&A_1,\text{ }B_1^2=B_1,\text{ }A_1B_1=0,  \nonumber \\
A_2^2 &=&B_2^2=I,\text{ }A_2B_2=\Pi _{2.}  \label{eq:qqq2}
\end{eqnarray}
From Eqs. (\ref{eq:c}) and (\ref{eq:qqq}-\ref{eq:qqq2}), the concurrence is
obtained as

\begin{eqnarray}
&&{\cal C}(U_{d_1\times d_2})  \nonumber \\
&=&\frac 2{d_1d_2}\sqrt{\text{tr}(A_1)\text{tr}(B_1)}\sqrt{d_2^2-|\text{tr}%
(B_2)|^2}  \nonumber \\
&=&\sqrt{\left( 1-\frac{|\text{tr}(\Pi _1)|^2}{d_1^2}\right) \left( 1-\frac{|%
\text{tr}(\Pi _2)|^2}{d_2^2}\right) }  \nonumber \\
&=&\sqrt{\left( 1-\frac{1-(-1)^{d_1}}{2d_1^2}\right) \left( 1-\frac{%
1-(-1)^{d_2}}{2d_2^2}\right) }.
\end{eqnarray}
where we have used the identity $\text{tr}(\Pi _i)=\frac 12[1-(-1)^{d_i}]$.
We see that $U_{d_1\times d_2}$ is a maximally entangled operator for even $%
d_1$ and even $d_2.\,$The concurrence ${\cal C}(U_{d_1\times d_2})$ becomes $%
\sqrt{d_2^2-1}/d_2$ for even $d_1$ and odd $d_2,\sqrt{d_1^2-1}/d_1$ for odd $%
d_1$ and even $d_2,$ and $\sqrt{(d_1^2-1)(d_2^2-1)}/(d_1d_2)$ for odd $d_1$
and odd $d_2.$ In the limit of $d_i\rightarrow \infty ,$ the operator $%
U_{d_1\times d_2}$ is a maximally entangled operator.

\subsection{Controlled$^N$-NOT}
Now we consider the entanglement of quantum gates. Let us
see the controlled$^N$-NOT gate\cite{CNN} which includes the controlled-NOT
gate\cite{CN} ($N=1$) as a special case. It is defined as

\begin{eqnarray}
\text{{\bf C}}^N\text{{\bf -NOT}} &=&e^{-i{\frac \pi {2^{N+1}}}(1-\sigma
_z)^{\otimes N}\otimes (1-\sigma _x)}  \nonumber \\
&=&I^{\otimes N}-2P_z^{\otimes N}\otimes P_x  \nonumber \\
&=&I^{\otimes k}\otimes I^{\otimes N+1-k}\nonumber\\
&&-2P_z^{\otimes k}\otimes
P_z^{\otimes N-k}\otimes P_x,
\end{eqnarray}
where $P_\alpha =(1-\sigma _\alpha )/2$ $(\alpha =x,y,z)$ are the projectors
satisfying $P_\alpha ^2=P_\alpha .$ The above equation implies that the ($%
N+1)$-th qubit flips if and only if all the other $N$ qubits is in the state 
$|1\rangle ^{\otimes N}.$ We split the whole system into two subsystems, one 
$2^k$-dimensional subsystem and another $2^{N+1-k}$-dimensional subsystem.
So in fact we are
studying the entanglement of an operator on $2^k\times 2^{N+1-k}$ systems.
Using the identity tr$(P_\alpha )=1$ and Eq.(\ref{eq:c}), the concurrence is
obtained as

\begin{equation}
{\cal C}=2^{1-N}\sqrt{(2^k-1)(2^{N+1-k}-1)}.
\end{equation}
For $N=1,k=1$, the concurrence ${\cal C}=1.$ So the controlled-NOT gate is a
maximally entangled operator. It is interesting to see that we can use this
maximally entangled gate to generate a maximally entangled two-qubit state.
As the local unitary operation does not change the amount of entanglement, we
know that the concurrence for controlled-NOT gate is the same as the unitary
operator $e^{-i{\frac \pi 4}\sigma _z\otimes \sigma _x},$ whose concurrence
is also 1.

\subsection{$\sigma _z^{\otimes N}$}

As a final example we investigate the following operator
which can be generated by the many-body Hamiltonian $\sigma _z^{\otimes N},$

\begin{equation}
V(\theta)=e^{-i\theta \sigma _z^{\otimes N}}=\cos \theta \text{ }I^{\otimes N}-i\sin\theta \sigma _z^{\otimes N}\text{ }.  \label{eq:ppp}
\end{equation}
The operator $V(\pi/4)$ can be used to generate GHZ state.
Like the above discussions we split the whole system into two subsystems,
one $2^k$-dimensional subsystem contains $k$ $(1<k<N)$ parties and another $%
2^{N-k}$-dimensional subsystem contains $N-k$ parties. We have 
\begin{eqnarray}
\mu  &=&\cos \theta \text{ },\nu =-i\sin \theta \text{ },  \nonumber \\
A_1 &=&I^{\otimes k},A_2=I^{\otimes N-k},  \nonumber \\
B_1 &=&\sigma _z^{\otimes k},B_2=\sigma _z^{\otimes N-k}.
\end{eqnarray}
This unitary operator belongs to the orthogonal case and the operators $A_i$
and $B_i$ are obvious self-inverse, so the concurrence is simply given by $%
{\cal C}=2|\mu \nu |=|\sin (2\theta )|,$ which clearly displays the
disentangled (maximally entangled) character of the unitary operator for $%
\theta =0,\pi /2(\theta =\pi /4).$

\section{conclusions}
We have studied the entanglement of unitary  operators acting on the state-space
of  $d_1\times d_2$ quantum systems.
We have  used as  entanglement measures, linear entropy and
concurrence. The former measure allows to make an explicit connection
(Eq.(\ref{eq:relation}) between the entanglement of an operator
and its entangling power.
Whereas the latter measure has been exploited in order to study
the bipartite entanglement of  a class of interesting unitary operators.
The existence of  some, more-or-less a direct, relation between operator
 entanglement quantified by concurrence and entangling capabilities
is an open issue.

\acknowledgments
This work has been supported by the European Community through grant
IST-1999-10596 (Q-ACTA).

\begin{figure}
\begin{center}
\epsfig{width=10cm,file=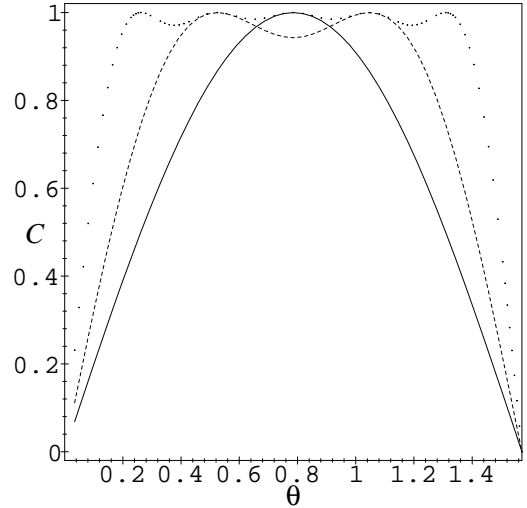}
\caption[]{The concurrence against $\theta$ for different $j$: $j=1/2$ (solid line), $j=1$ (dashed line), $j=5/2$ (dotted line).} 
\end{center}
\end{figure}

\end{document}